\documentclass[review]{elsarticle}

\usepackage{lineno,hyperref}
\modulolinenumbers[5]

\journal{Journal of \LaTeX\ Templates}

\begin{document}

\begin{frontmatter}

\title{Detection mechanism in highly sensitive ZnO nanowires network gas sensors}

\author[LISTaddress]{Nohora Caicedo\fnref{equalcontrib}}
\ead{noriliz@gmail.com}

\author[LISTaddress]{Renaud Leturcq\fnref{equalcontrib}\corref{mycorrespondingauthor}}
\cortext[mycorrespondingauthor]{Corresponding author}
\ead{renaud.leturcq@list.lu}
\fntext[equalcontrib]{Contributed equally to this work}

\author[ICTEAMaddress]{Jean-Pierre Raskin}
\ead{jean-pierre.raskin@uclouvain.be}

\author[ICTEAMaddress]{Denis Flandre}
\ead{denis.flandre@uclouvain.be}

\author[LISTaddress]{Damien Lenoble}
\ead{damien.lenoble@list.lu}



\address[LISTaddress]{Material Research and Technology Department (MRT), Luxembourg Institute of Science and Technology (LIST), Belvaux, Luxembourg}
\address[ICTEAMaddress]{Institute of Information and Communication Technologies, Electronics and Applied Mathematics (ICTEAM), Université catholique de Louvain (UCLouvain), Louvain-la-Neuve, Belgium}

\begin{abstract}
Metal-oxide nanowires are showing a great interest in the domain of gas sensing due to their large response even at a low temperature, enabling low-power gas sensors. However their response is still not fully understood, and mainly restricted to the linear response regime, which limits the design of appropriate sensors for specific applications. Here we analyse the non-linear response of a sensor based on ZnO nanowires network, both as a function of the device geometry and as a response to oxygen exposure. Using an appropriate model, we disentangle the contribution of the nanowire resistance and of the junctions between nanowires in the network. The applied model shows a very good consistency with the experimental data, allowing us to demonstrate that the response to oxygen at room temperature is dominated by the barrier potential at low bias voltage, and that the nanowire resistance starts to play a role at higher bias voltage. This analysis allows us to find the appropriate device geometry and working point in order to optimize the sensitivity. Such analysis is important for providing design rules, not only for sensing devices, but also for applications in electronics and opto-electronics using nanostructures networks with different materials and geometries.
\end{abstract}


\end{frontmatter}


\section{Introduction}

Electronic devices based on nanowires networks are showing a growing interest in the field of weareable electronics due to their low-cost and low-temperature processing, compatible with polymer-based substrates, and their mechanical flexibility, compatible with flexible substrates. A wide variety of materials have been used and interesting electronic properties have been achieved with nanowires networks, opening to an electronic architecture based only on nanowires networks. Silver nanowires or carbon nanotubes have been demonstrated as to be highly conductive transparent and flexible electrodes in organic transparent electronics \cite{LangleyD2013a,YeShengrong2014a,SannicoloT2016a,DuJinhong01,HuLiangbing2010a}. Carbon nanotubes can be used in active electronic devices \cite{HuLiangbing2010a}. Metal oxides nanowires, such as ZnO or SnO$_2$, have been studied for active devices \cite{SunB01,KoSH01,Unalan01} and sensors \cite{ChenXianping2013a}. In the field of gas sensors, metal oxide nanostructures have further proved to be highly sensitive at room temperature \cite{ZhangJun2016a} due to their large surface over volume ratio, as compared with the commercially used thin films, which usually work at high temperatures (typically above 200 $^\circ$C). This advantage opens new opportunities for low power gas sensors that will be required in the development of sensor networks for the Internet of Things \cite{PenzaM2014a}. However, the sensing mechanism at room temperature still remains not fully explained by current models for metal oxide based gas sensors.

One of the main difficulty in using devices based on nanowires networks lies in their complex electrical properties, due to the interplay between the conduction through individual nanowires and through the nanowire-to-nanowire junctions, in addition to the percolating conduction paths for networks with low density. Several models have been developed for highly conductive metallic nanowires networks, such as carbon nanotubes \cite{YanagiK2010a} or silver nanowires \cite{Bellew2015a}, often neglecting the contribution of the network. In the case of semiconducting nanowires, the junctions are often of great importance, as demonstrated in Figs. 1a and 1b. The interface states at the junction lead to band bending in the semiconductor close to the junction. This band bending creates a strong change of the local resistance, and to non-linear current-voltage ($I-V$) characteristics. In the case of gas sensors, an important question in order to understand the response to gases is the influence of the core nanowire vs. the junction on the electrical conduction.  Non-linear transport models have been developed for granular metal oxide films, in particular in the context of ZnO varistors \cite{BernasconiJ1977a,MahanGD1979a,SedkyA2007a,SedkyA2012a}. However, within these models, the response to gas was restricted to the linear regime, i.e. on the conductance \cite{Barsan2001a,YamazoeN2008a}. The full analysis of the non-linear transport regime is commonly performed for Schottky diode gas sensor \cite{ShivaramanMS1976a,PotjeKamloth2008a} but, to our knowledge, was never investigated for devices based on nanowires networks.

\begin{figure}
	\includegraphics[width=\columnwidth]{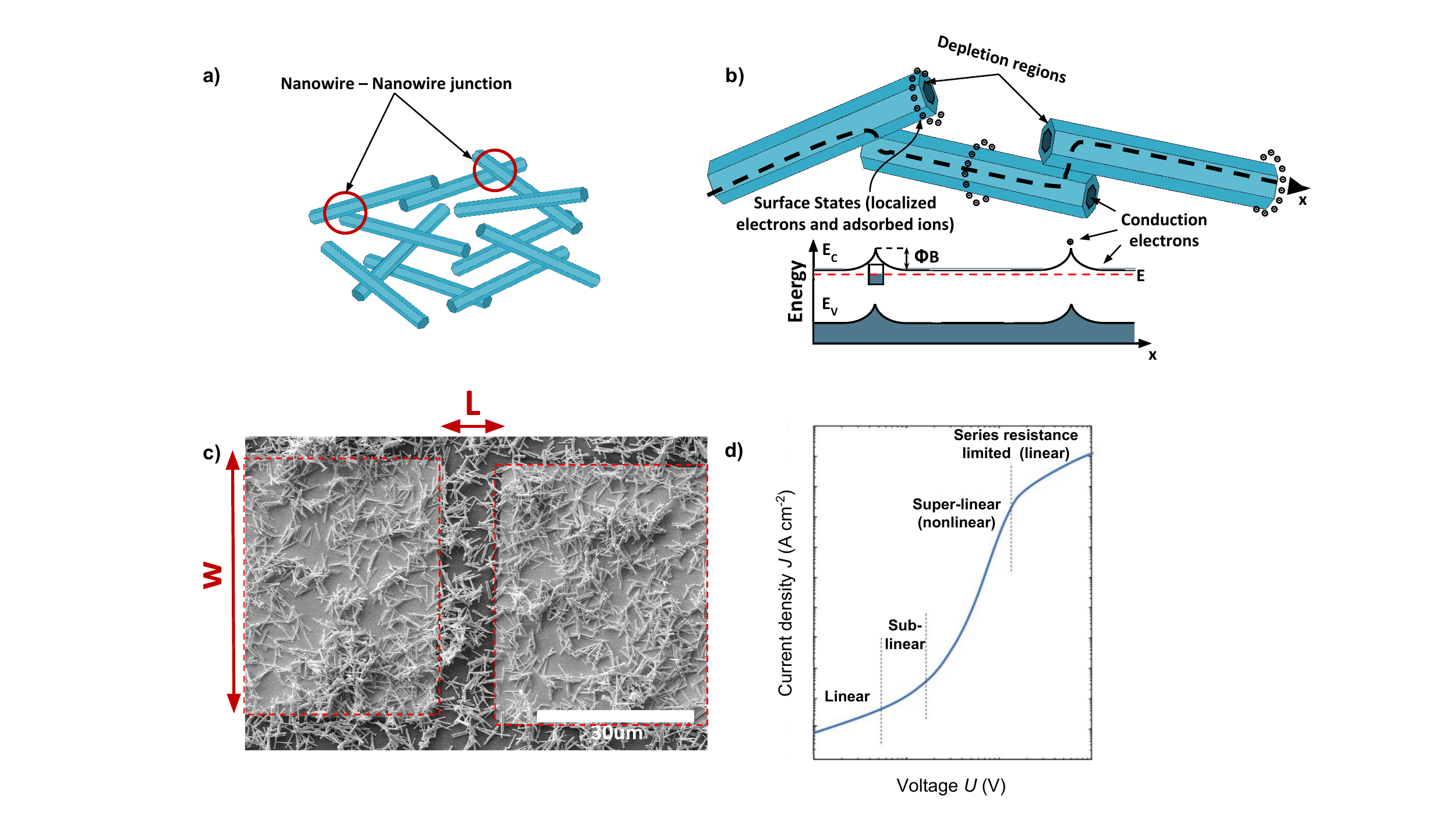}
	\caption{Nanowires electrical transport behavior. (a) Scheme of the nanowires network-based film. (b) Schematic cross sectional view of network-based sensing film with electronic-energy bands representation of semiconductors in equilibrium, where $E_C$ and $E_V$ are the conduction and valence band energies, $\Phi_B$ the barrier height and $E_F$ the Fermi level. (c) Scanning electron micrograph of a typical device with ZnO nanowires deposited on pre-fabricated electrodes (light gray), where $L$ is the gap between the electrodes and $W$ is the contacts width. (d) Expected current density vs. voltage ($j-V$) non-linear characteristic behavior metal oxide nanowires network, plotted from Eq.~\ref{eqn:4}.}
	\label{fig:Scheme_and_model}
\end{figure}

Here we extend the models of granular metal oxide films to the investigation of ZnO nanowires network gas sensor. By an analysis of devices with several geometries and measurements performed at different temperatures, we disentangle the effect of the junctions and the series resistance, coming either from the conduction along the nanowires or to the contacts. This allows us to validate the model, and extract the physical parameters of the sensor. 
At last we use the model in order to explain the very high sensitivity of these devices to oxygen exposure at room temperature. Oxygen is used as prototype oxidizing gas, allowing us to extract the direct dependence of the microscopic parameters on the gas concentration.  This approach allows us to choose the appropriate working point of the device. The analysis is important not only for gas sensors, but more generally for devices based on nanowires networks, such as junction-based devices or transistors, in order to tailor their electrical behavior.

\section{Experimental}

	\subsection{Sample fabrication}

ZnO nanowires were obtained by liquid phase synthesis from Zinc chloride (ZnCl$_2$) and hexamethylenetetramine (HMTA) precursors, as described elsewhere \cite{CaicedoN2016a}. All chemicals were analytical grade reagents (from Sigma Aldrich) and used as reactants without further purification. The amounts of ZnCl$_2$ and HMTA are equimolar in all our experiments corresponding to 10 mM. The precursors are first prepared as separate solutions of 1 M concentration. The reaction solution was prepared by mixing 1 mL of these concentrated solutions with 100 mL of MilliQ water (18.2 M$\Omega$.cm) in separate containers making the reaction solution at 10 mM. Subsequently, precursors are completely dissolved in water and stored in a 250 mL round glass vessel which fits into a Radleys Tech Carrousel able to heat and stir simultaneously under argon condition at 85$^\circ$C for 100 min. The magnetic stirring has always been controlled and kept constant during the synthesis in order to avoid agglomerates. The resulting solution is filtered, and the precipitate is washed with MilliQ water several times and kept in ethanol.  In this investigation, we have chosen nanowires obtained after one growth cycle (see Ref.~\citenum{CaicedoN2016a}) with a length of $2.5 \pm 0.3$ $\mu$m and a diameter of $100 \pm 40$ nm.

The devices were fabricated on a Si wafer with 270 nm thermal SiO$_2$. Arrays of electrodes with various geometries are realized by optical lithography, e-beam evaporation of Ti (5 nm) and Au (50 nm) and lift-off. For the investigation of the device geometry, we have used rectangular electrodes (see Fig.\ref{fig:Scheme_and_model}(c)) with various contact gaps ($L$) of 2, 5, 10 and 20 $\mu$m, and various contact 
widths ($W$) of 10, 50, 100, 200 and 300 $\mu$m. These small contacts are connected to larger 300 $\mu$m square pads for probing. For the gas sensor measurements, we have used interdigitated electrodes with a gap ($L$) of 20 $\mu$m and a total 
width ($W$) of 1050 $\mu$m. Once the synthesized nanowires are cleaned and rinsed, they are kept in ethanol solution. This solution is placed under sonication to enable a well-dispersed solution. The ZnO nanowire solution was drop-casted (1 droplet of 100 $\mu$L) on the pre-defined contacts and the solvent is dried in air at room temperature. The morphology of the samples was examined by scanning electron microscopy (SEM, HELIOS Nanolab 650).

	\subsection{Electrical measurements and gas sensing experiments}

$I-V$ characteristics of the ZnO nanowires network devices were measured first in vacuum using a Field-Upgradeable Cryogenic Probe Station CPX Model from Lake Shore Cryotronics, Inc. The CPX was operated in the temperature range from 100 K to 400 K by using liquid nitrogen as a continuous refrigeration system and a controlled heater on the sample stage. For gas sensing, a fixed flow of oxygen and nitrogen was introduced, with a ratio controlled by mass flow controllers. A Keithley 4200 source measure unit was used to record the $I-V$ characteristics of the nanowire-based network.

\section{Results and discussion}

	\subsection{Oxygen sensing results}

\begin{figure}
	\includegraphics[width=0.6\columnwidth]{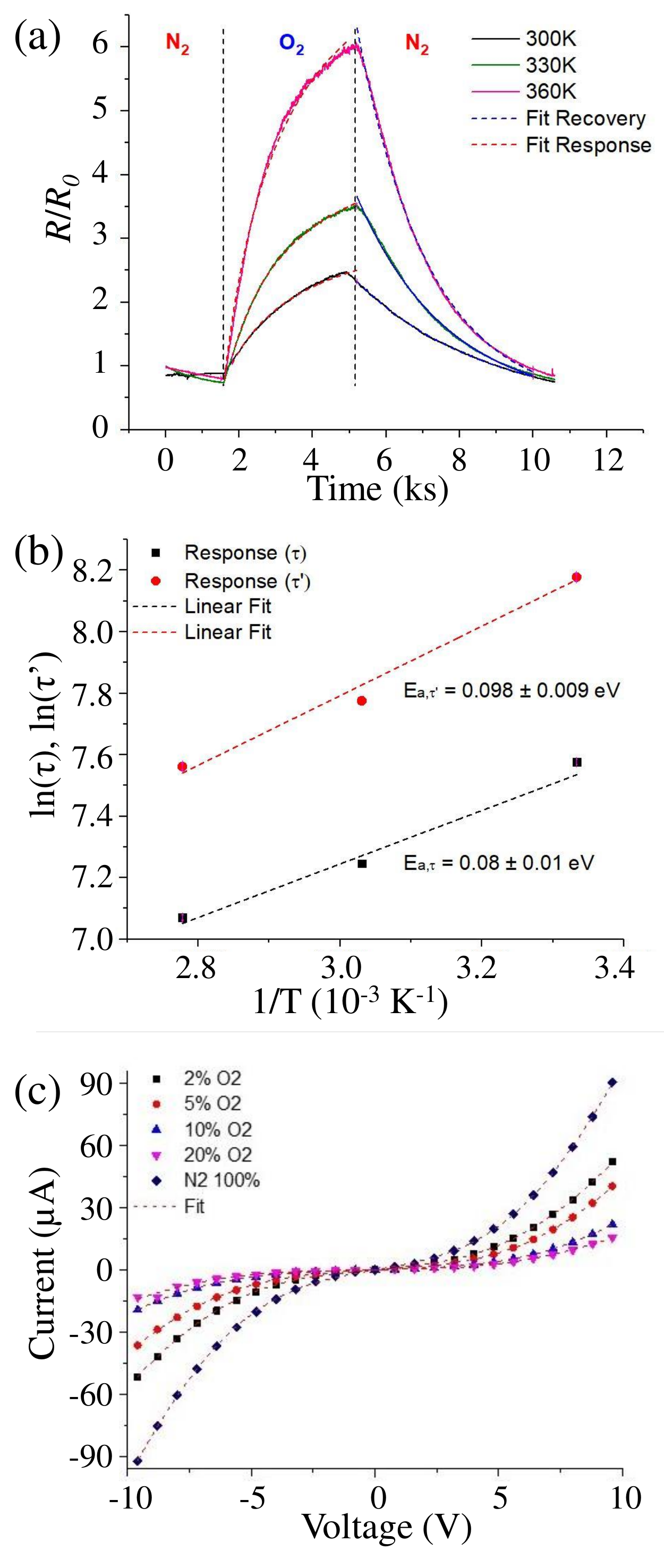}
	\caption{(a) Dynamic sensing characteristics of the ZnO nanowires network sensor to 5\% concentration of oxygen for different temperatures, with the voltage fixed at 5 V. $t$ and $t'$ are the characteristic response and recovery times calculated by fitting the data. The device has 20 $\mu$m gap between the interdigitated electrodes of 1050 $\mu$m contact width. (b) Arrhenius plot of dynamic sensing characteristics of the device under 5\% concentration of oxygen for different temperatures. The activation energy can be calculated from the slope $= E_a/K_B$ of the graph. (c) $I-V$ characteristics under different oxygen concentrations at room temperature. The experimental points have been fitted by the non-linear conduction model (see Eq.~\ref{eqn:4}).}
	\label{fig:Sensing_properties}
\end{figure}

Here we use oxygen as model gas since it is a prototype oxidizing gas \cite{KohlD1989a} and we can directly relate the amount of adsorbed species to the oxygen concentration (or partial pressure). In order to minimize the series resistance, for the gas sensing experiments we use interdigitated electrodes with a gap $L = 20$ $\mu$m and a total contact 
width $W = 1050$ $\mu$m. The influence of the oxygen concentration on the resistance is presented in Fig.~\ref{fig:Sensing_properties}(a) at various temperatures from room temperature to 360 K, showing that the sensor resistance increases when exposed to oxygen molecules as compared to nitrogen. This response is typical to oxidizing gases. In this case oxygen molecules are adsorbed on the surface, and form adsorbed $O_2^{-}$ species, which are the main stable species at a temperature below 100 $^\circ$C \cite{YamazoeN1979a}, by extracting one electron from ZnO. This leads to a decrease of the carrier concentration and thus an increase of the resistance of the device. We first note here the very high response (up to 150 \%) of the device even at room temperature for 5\% oxygen concentration. Similar high sensitivity was already reported for oxygen sensing using ZnO nanostructures \cite{AhmedF2013a}, and many other gases have been successfully detected at room temperature by ZnO nanostructures \cite{ZhuLing2017a}.

Fig.~\ref{fig:Sensing_properties}(b) shows the response and recovery times of the sensor as a function of the temperature, as well as an Ahrrenius fit of the data. The extracted activation energy for the response and recovery are, respectively, $98 \pm 9$ meV and $80 \pm 10$ meV, which are low values as compared with usual ZnO gas sensors ranging from 0.6 to 1.1 eV \cite{WatanabeHideo1965a,ZhangPeishuo2014a}. This is another indication of the very interesting response of nanowires network sensors for room-temperature operation. The main question arising from nanowires network sensors and that was never fully elucidated is whether the decrease in carrier concentration influences predominantly the resistance of the nanowires or the barrier height at the nanowire junctions. In order to answer this question, we have measured the non-linear $I-V$ curves of the devices for various values of the oxygen concentration, as shown in Fig.~\ref{fig:Sensing_properties}(c), with the aim at extracting the quantities governing the microscopic conduction mechanism. In the following parts of the article we investigate in details the model that will be applied on the nanowires network device for fully understanding this non-linear behaviour.

	\subsection{Conduction model in nanowires network}

The model that we use for metal oxide nanowires network is based on the schematic view depicted in Fig.~\ref{fig:Scheme_and_model}(b). We assume that the main effect of the junctions between nanowires is to locally change the potential seen by the conduction electrons, thus creating potential barriers with height $\Phi_B$. We assume a depletion of the surface, which is expected for ZnO surface, leading to a barrier potential for electrons. The current through the junctions can then be taken into account using the usual thermionic emission theory \cite{SedkyA2007a}, with possible influence of tunneling current in the case of narrow junctions. In addition to the barriers created at the junctions, nanowires have themselves a finite resistance, which is taken into account as a series resistance between each junction.

Using a similar approach than for granular thin films \cite{VarpulaA2010b,VarpulaA2010a}, the Kirchhoff law is applied is order to relate the total voltage across the network, $U$, and the voltage drop at a junction, $U_j$:
\begin{equation}
U = N_j U_j + R_s I \mathrm{ ,} \label{eqn:1}
\end{equation}
where $R_s$ is the effective series resistance, taking into account both the longitudinal resistance of nanowires, the length of current path and the contact resistances, and $I$ is the total current in the device. $N_j$ is the average effective number of junctions along the current path. Here we assume that all junctions and nanowires have uniform properties, and we can then assume that neither $N_j$ nor $R_s$ depends on the voltage. A model assuming a distribution of electrical properties would require a self-consistent calculation of the percolation paths, which would not lead to an analytical model, and goes beyond our approach.

The current across the junctions is assumed to be dominated by thermionic emission in the investigated temperature range, leading to the following current-voltage dependence:
\begin{equation}
I(U)=I_1 \left( e^\frac{q(U-R_s I)}{n_1 N_{j,1} k_B T} - 1 \right) - I_2 \left( e^\frac{q(U-R_s I)}{n_2 N_{j,2} k_B T} - 1 \right) \mathrm{ ,} \label{eqn:2}
\end{equation}
where the two terms correspond to the two-sided junctions that lead to a symmetric non-linear behavior, which are assumed to have different values for fitting purposes as a confirmation of the symmetric behavior. $q$ is the absolute electron charge, $n$ is the ideality factor of the junctions, $T$ the absolute temperature and $k_B$ the Boltzman constant. The values $I_1$ and $I_2$ are related to the temperature and the microscopic properties of the junctions, i.e. the barrier height $\phi_B$ and the Richardson constant $A$:
\begin{equation}
I_{1(2)}=A_{1(2)} T^2 e^{-\frac{\phi_B}{k_B T}} \label{eqn:3}
\end{equation}
The $I-V$ curve resulting from Eq.~\ref{eqn:2} is depicted schematically in Fig.~\ref{fig:Scheme_and_model}d, where different regimes can be distinguished \cite{VarpulaA2010b}. At low bias voltage, the current is limited by the potential barriers created by the grain-to-grain boundaries, leading to a linear dependence with high resistance \cite{MahanGD1983}. At higher voltage, thermionic emission above or tunneling of carriers through the potential barrier leads to a sublinear dependence \cite{LevinsonLM1975a,MahanGD1979a}. At even higher voltage, the barrier height decreases when increasing the voltage and the current increases exponentially, leading to the super-linear region. At the highest voltage, the series resistance limits the current and the $I-V$ curve returns to a linear behaviour \cite{LevinsonLM1978a,VarpulaA2008a,VarpulaA2010b,VarpulaA2010a,VarpulaA2011a}.

The $I-V$ characteristics in vacuum were measured in the bias range from -20 V to 20 V for multiple devices with different geometries. In order to fit the experimental data, we solve separately each branch of Eq.~\ref{eqn:2} by using the Lambert W-function:
\begin{equation}
I(U) = \frac{n_{1(2)} N_{j,1(2)} k_B}{q R_s} \mathcal{W} \left( \frac{qI_{1(2)} R_s}{n_{1(2)} N_{j,1(2)} k_B} e^{\frac{q(U+R_s I_{1(2)}) }{n_{1(2)} N_{j,1(2)} k_B }} \right) \mathrm{ .} \label{eqn:4}
\end{equation}
The fitting of the experimental data with Eq.~\ref{eqn:4} allows us to determine the parameters $R_s$, $n_1 N_{j,1}$ and $n_2 N_{j,2}$ , $I_1$ and $I_2$.

\begin{figure}
	\includegraphics[width=\columnwidth]{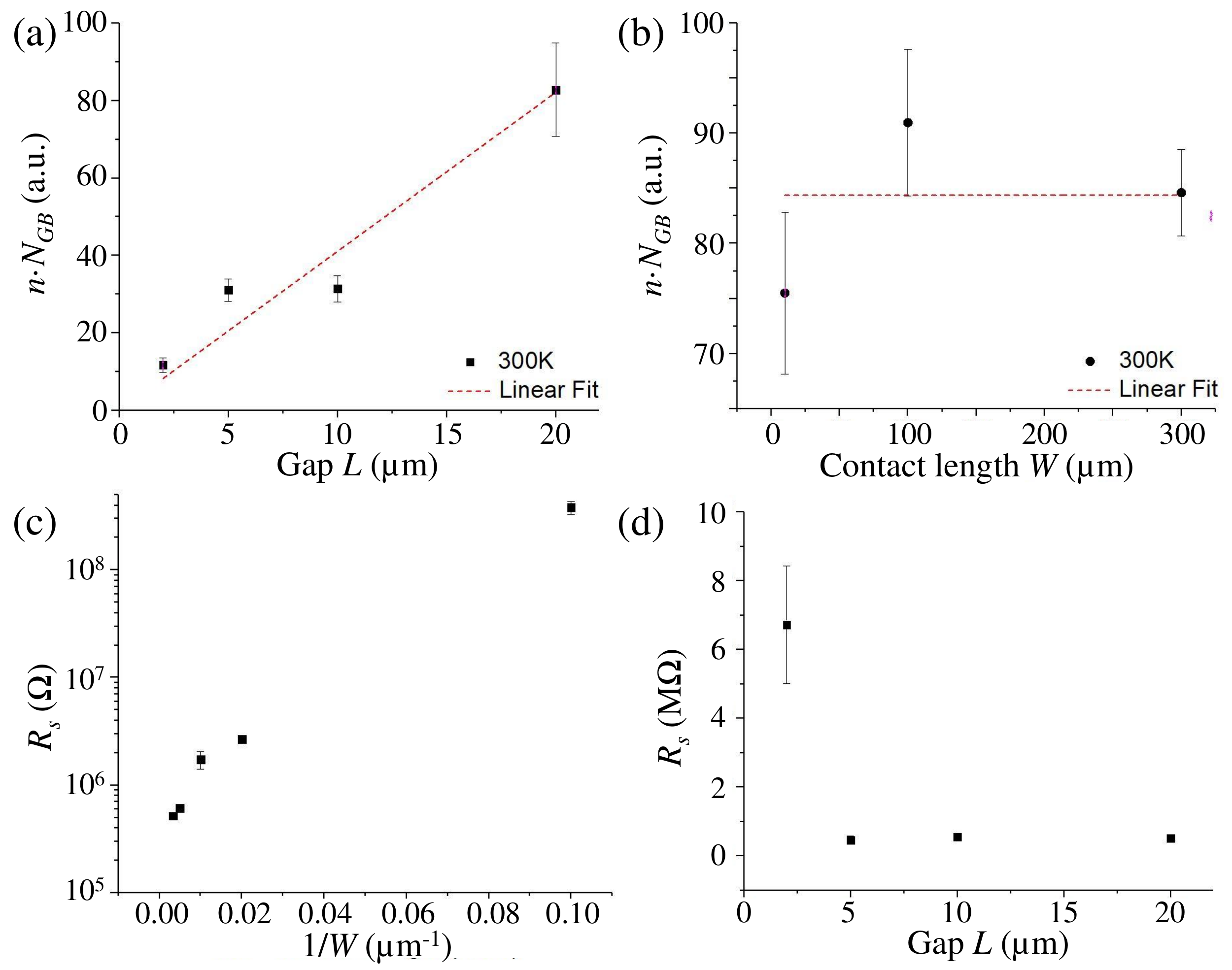}
	\caption{Dependence of the parameter $n N_i$ ($n$ being the ideality factor, $N_j$ the average number of junctions along the current path) and the series resistance $R_s$ on the contact geometrical parameters $L$ and $W$ for samples with as-deposited nanowires . (a) $n N_j$ as a function of the gap $L$ for $W = 10$ $\mu$m (the dashed line is a linear fit). (b) $n N_j$ as a function of the contact 
		width $W$ for $L = 20$ $\mu$m. (c) $R_s$ as a function of the inverse of the contact 
		width $W$ for $L = 20$ $\mu$m. (d) $R_s$ as a function of the gap $L$ for $W = 10$ $\mu$m. The error bars correspond to the standard deviation of the mean generated by the fit for each value.}
	\label{fig:nN_and_Rs_vs_geometry}
\end{figure}

In order to confirm the validity of the model, we have determined the parameters $R_s$ and $n N_j$  for a set of devices with variable geometry (i.e. various contact gap lengths $L$ and contact 
widths $W$) and a constant nanowire density. The variations of the fitting parameters as a function of $L$ and $W$ for as-deposited nanowires are shown in Fig.~\ref{fig:nN_and_Rs_vs_geometry}. The most important result is the linear dependence of $n N_j$ as a function of the gap between contacts, $L$ (Fig.~\ref{fig:nN_and_Rs_vs_geometry}(a)), while it is almost independent on the contact 
width, $W$ (Fig.\ref{fig:nN_and_Rs_vs_geometry}(b)). Assuming that the ideality factor of the junctions does not change from device to device, these dependencies are logically understood as the dependence of the average effective number of junctions along the current path, $N_j$, which is expected to vary linearly as a function of the gap between electrodes, and not depend on $W$. This result is a strong validation of the hypothesis that the number of junctions participating to the transport is only given by the geometry of the device. Moreover the series resistance $R_s$ shows a linear dependence as a function of $1/W$, which is also expected from the geometry of the device where the resistances are combined in parallel along $W$. This further validates the hypothesis in using the Kirchoff’s law for combining individual resistive elements. The behavior of $R_s$ as a function of $L$ (Fig.~\ref{fig:nN_and_Rs_vs_geometry}(d)) is less obvious due to the combined effect of the contact resistance (independent on $L$) and the number of nanowires in series (linear in $L$). Moreover the gap of 2 $\mu$m is smaller than the nanowire length (3 $\mu$m), where the model is not valid anymore.

\begin{figure}
	\includegraphics[width=\columnwidth]{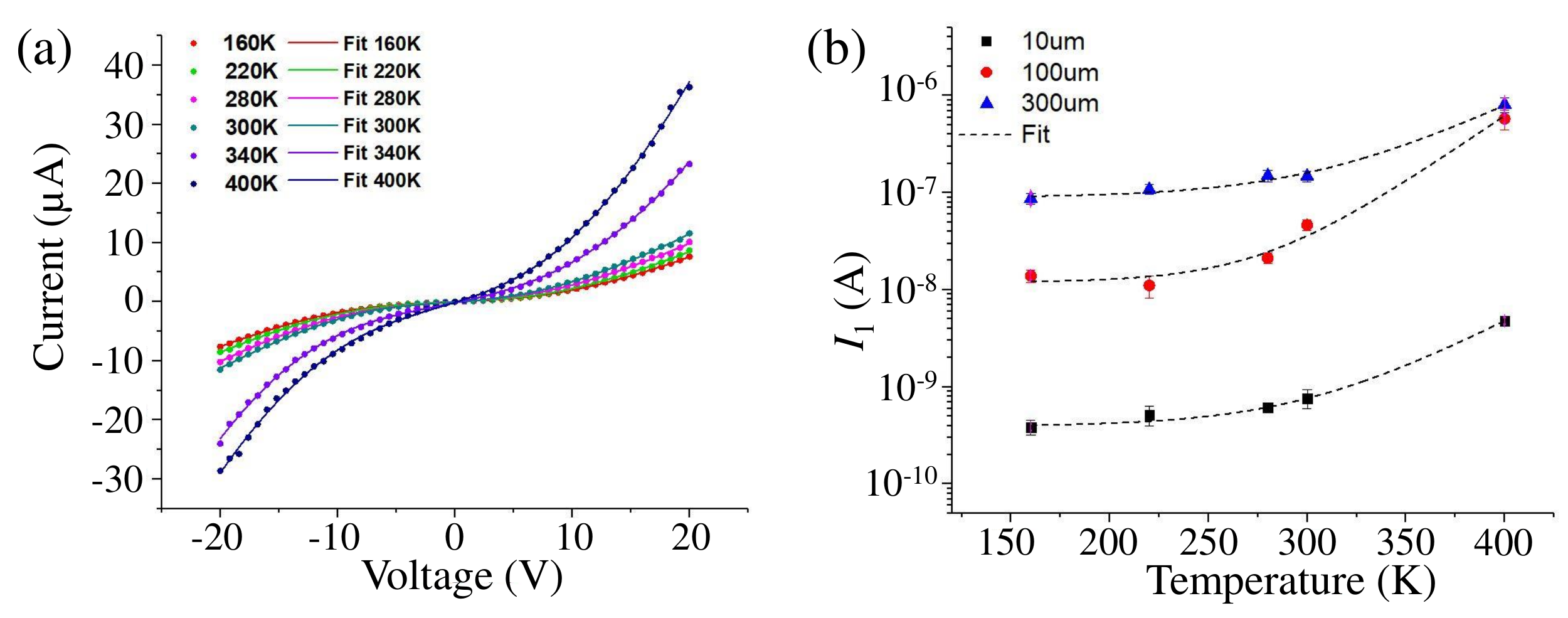}
	\caption{(a) Temperature dependence of the $I-V$ characteristics for ZnO nanowires network-based thin film under vacuum. The measured device corresponds to $L = 20$ $\mu$m and $W = 300$ $\mu$m. (b) $I_{1,2}$ extracted from the fit of the $I-V$ curve vs. the temperature. The error bars are standard errors of the fit parameters.}
	\label{fig:temperature-dependence}
\end{figure}

In order to extract the barrier height $\phi_B$ and the Richardson constant $A$, $I-V$ curves were measured at different temperatures from 160 to 400 K, as shown for a specific device in Fig.~\ref{fig:temperature-dependence}(a). For each temperature, we fit the $I-V$ characteristics with Eq.~\ref{eqn:4} in order to obtain the parameters $I_1$ and $I_2$. $I_1$ and $I_2$ are then plotted as a function of the absolute temperature, as shown in Fig.~\ref{fig:temperature-dependence}(b) for three specific devices, revealing a clear exponential dependence behaviour. From this result, a second fit with Eq.~\ref{eqn:3} is then performed in order to extract the parameters $A$ and $\phi_B$ for the different tested geometries.

\begin{figure}
	\includegraphics[width=\columnwidth]{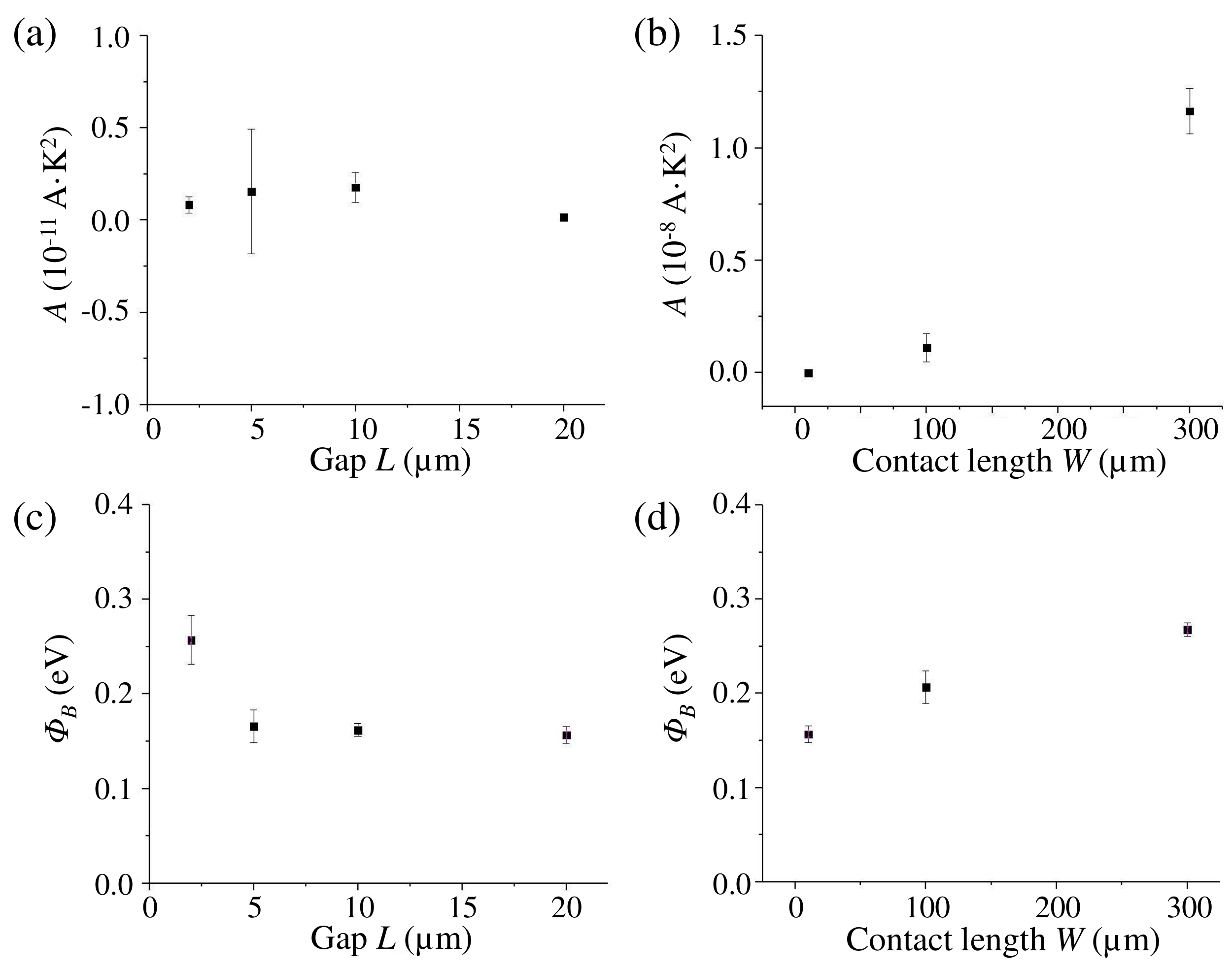}
	\caption{Dependence of the Richardson constant $A$ and the barrier height $\phi_B$ as a function of the geometrical parameters $L$ and $W$. (a) $A$ as a function of the gap $L$ for $W = 10$ $\mu$m. (b) $A$ as a function of the contact 
		width $W$ for $L = 20$ $\mu$m. (c) $\Phi_B$ as a function of the gap $L$ for $W = 10$ $\mu$m. (d) $\Phi_B$ as a function of the contact 
		width $W$ for $L = 20$ $\mu$m. The error bars correspond to the standard deviation of the mean generated by the fitting for each value.}
	\label{fig:A_and_PhiB_vs_geometry}
\end{figure}

The parameters $A$  and $\phi_B$ are plotted as a function of the device geometry parameters $W$ and $L$ in Fig.~\ref{fig:A_and_PhiB_vs_geometry}. $A$ mainly depends on the contact 
width $W$ and weakly on the gap $L$. This behaviour is expected since $A$ is the product of the junction surface area $S$ and of the normalized Richardson constant $A^\star = A/S$, which only depends on the effective mass of the semiconductor. The increase of $A$ as a function of $L$ is thus directly related to the increase of the global junction area due to the increase of the number of parallel junctions when the contact 
width increases. This increase is expected to be sub-linear for small $W$ due to current path flowing on the sides of the contacts, and becomes linear at high $W$, as observed experimentally. For the barrier height $\phi_B$, we first note that the obtained value ranges between 0.12 and 0.25 eV. These values are lower than the typical barrier height  of 0.7 - 0.9 eV observed at the grain boundaries of annealed ZnO \cite{MahanGD1979a,MantasPQ1995a}, for which the doping in the grains is as low as $10^{17}$ cm$^{-3}$. The low value can be explained by the high doping in ZnO nanowires due to intrinsic defects, as observed in ZnO thin film obtained by sol-gel, for which barrier height as low as 0.176 eV have been extracted \cite{YildizA2016a}. Figures \ref{fig:A_and_PhiB_vs_geometry}(c) and \ref{fig:A_and_PhiB_vs_geometry}(d) show that $\phi_B$ depends both on $W$ and $L$, which is surprising at first since the barrier height at each junction should not depend on the geometry of the network. The dependence in $L$ shows first a decrease from 2 to 5 $\mu$m, and then is almost independent on $L$ for spacing larger than 5 $\mu$m. We attribute this behaviour to the contribution of the Schottky barrier at the contact at small gap $L = 2$ $\mu$m since the nanowire length (3 $\mu$m in average) is then larger than this spacing. Above 5 $\mu$m, the barriers between nanowires dominate, and the extracted barrier height becomes independent on $L$, as expected. On the other hand $\phi_B$ decreases when $W$ becomes small. We can also attribute this behaviour to an artefact of the finite size of the device due to current paths flowing on the sides of the device for smaller values of $W$.

	\subsection{Application of the conduction model to gas sensing}

\begin{figure}
	\includegraphics[width=\columnwidth]{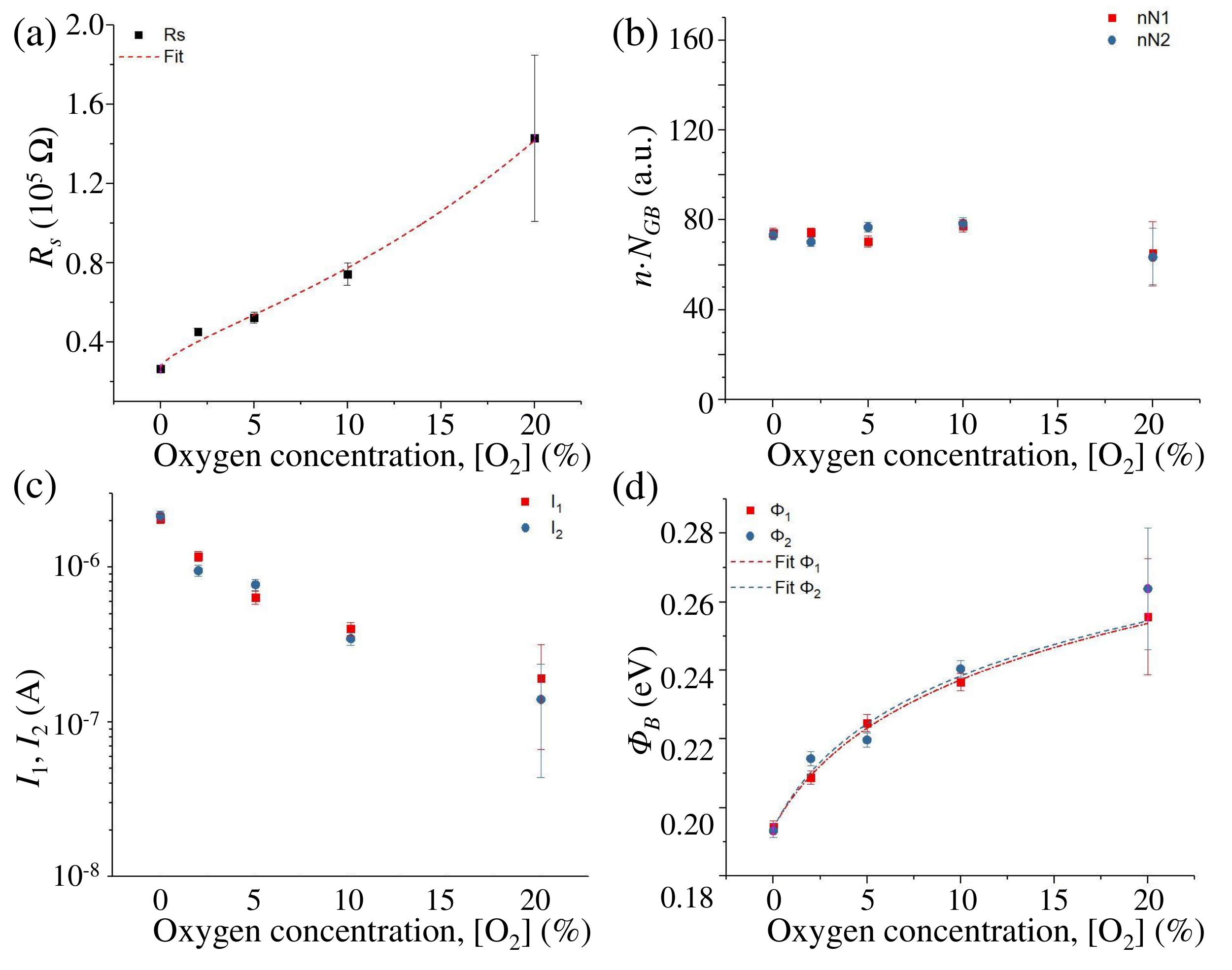}
	\caption{(a) Series resistance $R_s$, (b) parameter $n N_j$, (c) $I_1$ and $I_2$, and (d) $\Phi_B$ as a function of oxygen gas concentration. The measurements were performed at 300 K. The data in (a) and (d) are fitted, respectively, by Eqs.~\ref{eq:8} and \ref{eq:7} with the following parameters, $\Phi_{B,0} = 0.479 \pm 0.001$ eV, $C = 27 \pm 1$ meV and $p_0 = 2.6 \pm 0.1$ \% for $\Phi_1$, $\Phi_{B,0} = 0.480 \pm 0.004$ eV, $C = 27 \pm 1$ meV and $p_0 = 2.4 \pm 0.5$ \% for $\Phi_2$.}
	\label{fig:param_vs_o2_conc}
\end{figure}

We now show that the non-linear transport model described in previous part can be successfully applied to gas sensors based on nanowires networks. We analyze here the $I-V$ curves taken at different values of the oxygen concentration in Fig.~\ref{fig:Sensing_properties}(c). Each curve is fitted using the conduction model presented previously for nanowire-based networks. The extracted series resistance $R_s$ and the parameter $n N_j$ are plotted as a function of the gas concentration in Fig.~\ref{fig:param_vs_o2_conc}(a) and \ref{fig:param_vs_o2_conc}(b). $n N_j$ is almost independent on the gas concentration. The independence of $N_j$ is expected since the morphology of the nanowires network remains unchanged. The absence of variation of the ideality factor $n$ is also expected as long as the gas weakly influences the barrier height. We however point out recent studies on metal-insulator-SiC and graphene-Si diodes that have shown changes of the ideality factor under different environments or operating temperature \cite{NakagomiS2002a,KimHyeYoung2013a,LiuY2014a}. The absence of dependence in our case might be related to the nature of the ZnO-ZnO homojunction vs. the usually used heterojunction. The main variation due to the gas is first on the series resistance, which increases linearly as a function of the gas concentration. This dependence will be quantitatively analyzed with the model below.

The parameters $I_1$ and $I_2$ are plotted as a function of the oxygen gas concentration in Fig.~\ref{fig:param_vs_o2_conc}(c). These parameters show an almost exponential decrease when the gas concentration increases. In order to investigate the dependence on the barrier height $\phi_B$, we assume in Eq.~\ref{eqn:3} that the Richardson constant $A$ does not depend on the gas concentration. Indeed, $A$ is proportional to the junction area, and depends on the semiconductor effective mass \cite{Sze01}, which can be assumed independent on gas concentration for a given device and a given material. We thus assume for this device a constant $A$ extrapolated from the dependence demonstrated in Fig.~\ref{fig:A_and_PhiB_vs_geometry}(b), and we use Eq.~\ref{eqn:3} to extract $\phi_B$ as a function of the oxygen concentration, which is plotted in Fig.~\ref{fig:param_vs_o2_conc}(d). $\phi_B$ increases logarithmically as a function of the gas concentration.

In order to explain quantitatively the dependences of $\phi_B$ and $R_s$ as a function of the oxygen concentration (or partial pressure $p_{O_2}$), we have adapted existing gas sensor models to our current case of room-temperature sensing. We note that, at room temperature, the involved chemical reaction at the surface of ZnO is \cite{YamazoeN1979a}:
\begin{equation}
O_2 + e^{-} \leftrightarrow O_2^{-},ads \mathrm{ ,}
\end{equation}
with a reaction constant $K_{O_2}$ defined as
\begin{equation}
K_{O_2} =  \frac{[O_2^{-},ads]}{p_{O_2} [e]} \mathrm{ ,}
\end{equation}
where $[O_2^{-},ads]$ is the concentration of adsorbed species on the surface, $p_{O_2}$ the oxygen partial pressure, which is proportional to the oxygen percentage, and $[e]$ is the local concentration of electrons in ZnO. By approximating the barrier potential due to the depletion at ZnO surface by a square potential, with a width equals to the depletion length $w$ and height $\phi_B$, and assuming that the donors in ZnO with concentration $N_d$ are fully ionized in the barrier, the charge neutrality equation reads \cite{YamazoeN2008a}:
\begin{equation}
N_d w = [O_2^{-},ads] + \frac{Q_{ss}}{q} = K_{O_2} p_{O_2} [e] + N_{ss} \mathrm{ ,}
\end{equation}
where $Q_{ss}$ is the ionized surface charges when no oxygen is present and $N_{ss}$ the surface traps density. The local electron concentration is related to the barrier height through:
\begin{equation}
[e] = N_d e^{-\frac{\Phi_B}{k_B T}} \mathrm{ ,}
\end{equation}
and we can then write:
\begin{equation}
N_d w = K_{O_2} p_{O_2} N_d e^{-\frac{\Phi_B}{k_B T}} + N_{ss} \mathrm{ .} \label{eq:6.1}
\end{equation}

We note here that both the depletion length $w$ and the surface trap density $N_{ss}$ depend on the barrier height through the following equations \cite{Sze01}:
\begin{eqnarray}
q N_d w = \sqrt{ 2 \varepsilon_s N_d \left( \phi_B - k_B T \right) } \mathrm{ ,} \label{eq:6.2} \\
N_{ss} = D_{it} (E_G - \phi_0 - \phi_B + \phi_n) \mathrm{ ,}
\end{eqnarray}
where $\varepsilon_s$ is the permittivity of the semiconductor, $E_G$ its bandgap, $\phi_0$ the lower energy of the surface states and $\phi_n$ the difference between the Fermi level and the conduction band minimum. Due to these dependences, Eq.~\ref{eq:6.1} cannot be solved analytically. However these dependencies are weak as compared to the exponential dependence in Eq.~\ref{eq:6.1} and can be neglected as a first approximation, giving the following relationship between the barrier height and the oxygen partial pressure:
\begin{equation}
\frac{\phi_B}{k_B T} = \ln \left( K_{O_2} N_d \right) + \ln \left( \frac{p_{O_2}}{N_d w - N_{ss}} \right) \mathrm{ .} \label{eq:6.3}
\end{equation}
We note that this expression is not valid when $p_{O_2} \rightarrow 0$, since then $(N_d w - N_{ss}) \rightarrow 0$ and the ratio $\frac{p_{O_2}}{N_d w - N_{ss}}$ is undetermined. We however know that $\Phi_B$ remains finite when $p_{O_2} \rightarrow 0$, so that 
\begin{equation}
\lim_{p_{O_2} \rightarrow 0} \left( Ndw - N_{ss} \right) \propto p_{O_2} \qquad \mathrm{and} \qquad \lim_{p_{O_2} \rightarrow 0} \left( Ndw - N_{ss} \right) = C \mathrm{ .}
\end{equation}
For this purpose we assume the following simple dependence:
\begin{equation}
N_d w - N_{ss} \propto \frac{p_{O_2}}{p_{O_2} + C} \mathrm{ ,}
\end{equation}
and Eq.~\ref{eq:6.3} writes:
\begin{equation}
\Phi_B = A + B \ln \left( p_{O_2} + C \right) \mathrm{ ,} \label{eq:7}
\end{equation}
where $A$, $B = k_B T$ and $C$ are constants. As shown in Fig.~\ref{fig:param_vs_o2_conc}d, we obtain a very good fit of the data with Eq.~\ref{eq:7}, with the value of $B = 27 \pm 1$ meV very close to $k_B T = 25.9$ meV at 300 K, thus validating our assumptions and the proposed model.

The same model can also be applied for the resistance along ZnO nanowires. In the hypothesis where the nanowire diameter is much larger than the Debye length, $\lambda_D \ll D$, the change of the nanowire resistance is directly due to the change of effective nanowire diameter due to the depletion:
\begin{equation}
R \propto \left( \frac{D}{D-2w} \right)^2 \mathrm{ .} \label{eq:8}
\end{equation}
Using the function described above for expressing the barrier height as a function of the oxygen partial pressure with Eq.~\ref{eq:6.2} we deduce:
\begin{equation}
w = \frac{\sqrt{e \varepsilon_s N_d}}{q N_d} \sqrt{ A + B \ln \left( p_{O_2} + C \right) - B} \equiv \alpha \sqrt{ A + B \left( \ln \left( p_{O_2} + C \right) - 1 \right)} \mathrm{ ,}
\end{equation}
which we can use in Eq.~\ref{eq:8}. We have fitted the data of Fig.~\ref{fig:param_vs_o2_conc}(a) with the dependence of Eq.~\ref{eq:8} and the dependence of $\Phi_B$ deduced from Fig.~\ref{fig:param_vs_o2_conc}(d), showing that the dependence of $R_s$ can be successfully reproduced as well.

This analysis shows that this simple analytical model works well for ZnO nanowires network gas sensors. This is an important result which allows the direct analysis of the sensing behaviour of the device as well as the prediction of the sensitivity as described below.

	\subsection{Sensitivity of the nanowires network gas sensor}

\begin{figure}
	\includegraphics[width=\columnwidth]{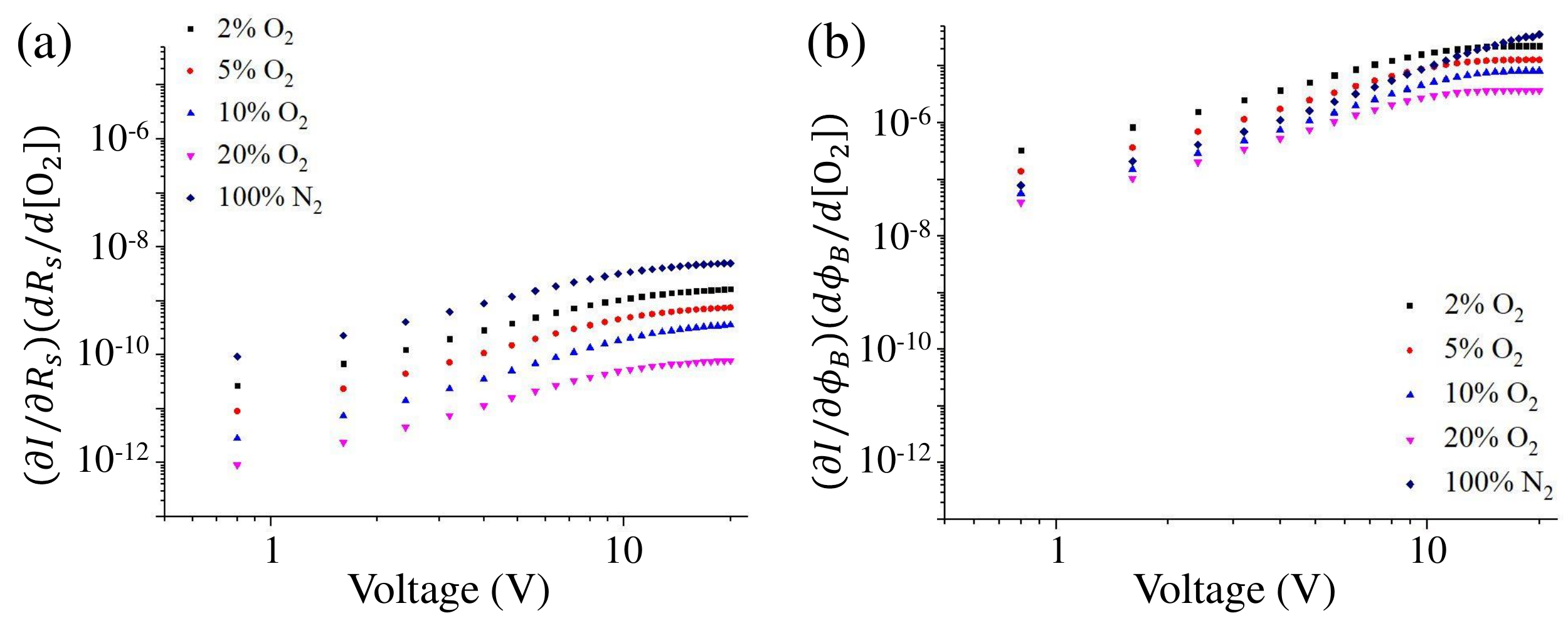}
	\caption{Respective contributions to the response $dI/d[O_2]$ due to (a) series resistance $R_s$ and (b) barrier potential $\phi_B$ as a function of the applied voltage. The measurements were performed at 300 K.}
	\label{fig:derivatives_contribution}
\end{figure}

At last, an important analysis in order to provide design rules for the sensor is to evaluate the relative contribution of the change in $R_s$ and the change in $\phi_B$ on the gas sensor response. We note that, from Eq.~\ref{eqn:4}, the global response of the current as a function of $O_2$ concentration [$O_2$], $dI/d$[$O_2$], at a given bias voltage $U$ can be split into:
\begin{equation}
dI/d[O_2]= \frac{\partial I}{\partial R_s} \frac{dR_s}{d[O_2]} + \frac{ \partial I}{ \partial \phi_B}  \frac{d\phi_B}{d[O_2]} \mathrm{ ,}	\label{eqn:5}
\end{equation}
where $dRs/d[O_2] = 28$ k$\Omega/$\% and $d \phi_B/d[O2] = 0.52-0.57$ eV/(2-20\%) are obtained from the fits in Fig.~\ref{fig:Sensing_properties}(c), and the partial derivatives are calculated numerically for each value of $[O_2]$ and $U$ from Eq.~\ref{eqn:4} using the parameters of Fig.~\ref{fig:param_vs_o2_conc}. In Fig.~\ref{fig:derivatives_contribution}, we plot both parts of Eq.~\ref{eqn:5} as a function the bias voltage $U$ and the oxygen concentration. Both contributions increase as a function of the bias voltage. We however note that the contribution due to the barrier height dominates the one of the series resistance for almost all values of oxygen concentration. Moreover, the response due to teh change of the barrier height is much more weakly dependent on the gas concentration as the series resistance, which is interesting when the linearity of the response is considered.

The large contribution of the barrier height on the sensitivity is expected to strongly depend on the device geometry. While analyzing the dependence of the series resistance and the barrier height as a function of the device geometry, we have shown that the contribution of the series resistance is proportional to $1/W$, while the contribution of the barrier height weakly depends on $W$. The choice of a large contact 
width $W$ for the final device is thus the main reason for the strong contribution of the barrier height. Interestingly the competitive contribution between the series resistance and the barrier height can be adjusted by varying the contact 
width and the gap between the contacts, thus giving a wide opportunity for adapting the device to targeted properties.

\section{Conclusion}

In conclusion, we demonstrate a new approach for analyzing the gas sensing response of metal oxide nanowires networks, by proposing a model where we can independently extract the contribution of the series resistance of the nanowires and the barrier height of the junctions between nanowires. By investigating devices with different geometries, we show that the measurements obtained on networks of ZnO nanowires match with a very good accuracy the expected equations. We further use this model to investigate the microscopic origin of oxygen sensing in this nanowires network based gas sensor. The influence of the gas on both the series resistance and the nanowire junctions are disentangled, demonstrating that the very high sensitivity at room temperature is related to the barrier height modulation. In the future, this new approach can be extended to other materials and other gases in order to tailor the response of sensing devices based on nanowires networks. Varying the geometry is an important factor that could change the conduction and sensing regime of nanowires networks, which could be used as an important tool for tayloring the response and/or performances beyond simple geometrical factors.

\section*{References}

\bibliography{biblio}

\end{document}